# Weak Antilocalization Effect up to ~ 120 K in the van der Waals Crystal $Fe_{5-x}GeTe_2$ with Near Room Temperature Ferromagnetism

*AUTHOR NAMES:* Zhengxian Li[1,†], Kui Huang[1,†], Deping Guo[2,†], Guodong Ma[3], Xiaolei Liu[1], Yueshen Wu[1], Jian Yuan[1], Zicheng Tao[1], Binbin Wang[1], Xia Wang[1,4], Zhiqiang Zou[1,4], Na Yu[1,4], Geliang Yu[3], Jiamin Xue[1], Jun Li[1,5]\*, Zhongkai Liu[1,5]\*, Wei Ji[2]\*, Yanfeng Guo[1,3,6,]\*.

AUTHOR ADDRESS

[1]School of Physical Science and Technology, ShanghaiTech University, Shanghai 201210, China

[2]Department of Physics and Beijing Key Laboratory of Opto-electronic Functional Materials & Micro-nano Devices, Renmin University of China, Beijing 100190, China

[3]National Laboratory of Solid State Microstructures and School of Physics, Nanjing University, Nanjing 210093, China

[4]Analytical Instrumentation Center, School of Physical Science and Technology, ShanghaiTech University, Shanghai 201210, China

[5]ShanghaiTech Laboratory for Topological Physics, Shanghai 201210, China

[6]State Key Laboratory of Surface Physics and Department of Physics, Fudan University, Shanghai 200433, China

[†]These authors contributed equally to this work.

*Email address: lijun3@shanghaitech.edu.cn, liuzhk@@shanghaitech.edu.cn, wji@ruc.edu.cn, guoyf@shanghaitech.edu.cn.

**ABSTRACT**: **The weak antilocalization (WAL) effect is known as a quantum correction to the classical conductivity, which never appeared in two-dimensional magnets. In this work, we reported the observation of a WAL effect in the van der Waals ferromagnet $Fe_{5-x}GeTe_2$ with a Curie temperature $T_c$ ~ 270 K, which can even reach as high as ~ 120 K. The WAL effect could be well described by the Hikami-Larkin-Nagaoka and Maekawa-Fukuyama theories in the presence of strong spin-orbit coupling (SOC). Moreover, A crossover from a peak to dip behavior around 60 K in both the magnetoresistance and magnetoconductance was observed, which could be ascribed to a rare example of temperature driven Lifshitz transition as indicated by the angle-resolved photoemission spectroscopy measurements and first principles calculations. The reflective magnetic circular dichroism measurements indicate a possible spin reorientation that kills the WAL effect above 120 K. Our findings present a rare example of WAL effect in two-dimensional ferromagnet and also a magnetotransport fingerprint of the strong SOC in $Fe_{5-x}GeTe_2$. The results would be instructive for understanding the interaction Hamiltonian for such high $T_c$ itinerant ferromagnetism as well as be helpful for the design of next-generation room temperature spintronic or twistronic devices.**

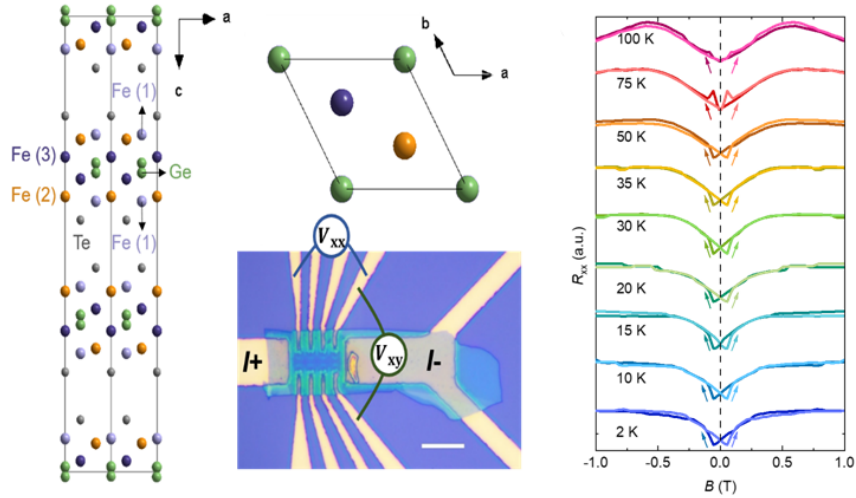

**KEYWORDS**: van der Waals ferromagnet, $Fe_{5-x}GeTe_2$, room-temperature Curie temperature, weak antilocalization effect, spin-orbit coupling.

The quasi-two-dimensional (2D) van der Waals (vdW) bonded magnets have captured vast attention due to their twofold merits for both fundamental study and technical applications. The vdW magnets in the 2D limit exhibits extraordinary properties that are scarcely observed in the bulk, arisen from the complex magneto-electrical, magneto-optical, or spin-lattice coupling effects[1-6]. Furthermore, their virtue of easy exfoliation into multi- or single layer makes them very convenient for constructing various novel heterostructures and devices, which would find potential applications in next-generation spintronic devices. In principle, in the isotropic 2D limit, the spontaneous symmetry breaking for the formation of intrinsic long-range magnetic order is generally prohibited by strong thermal fluctuations, which is known as the Merin-Wagner theorem[7]. Nevertheless, the long-range magnetic order has been realized in a broad range of vdW magnets, such as in mono- or few-layer $CrI_3$[8], $Cr_2Ge_2Te_6$[9], $Cr_2Si_2Te_6$[10], $VSe_2$[11], and $MnSe_2$[12], etc., owing to their intrinsic magnetocrystalline anisotropy which can open up a non-zero excitation gap in the low energy mode of the acoustic branch and hence stabilize the magnetic order against finite temperature.

For practical use, ferromagnetic (FM) semiconductors are always under intense pursue, especially those can work at room temperature. On the other side, high $T_c$ FM conductors that are desirable for building magnetic twistronic and spintronic devices are also very rare. In this regard, the Fe$_n$GeTe$_2$ ($n$ = 3, 4, 5) ferromagnets are very attractive due to the high $T_c$. The study on monolayer Fe$_3$GeTe$_2$ unveiled that though the $T_c$ is suppressed as compared with that of the bulk, it can be significantly enhanced even up to room temperature through ionic gating, thus offering opportunities for the usage in room temperature spintronic devices. Among the Fe$_n$GeTe$_2$ ($n$ = 3, 4, 5), Fe$_{5-x}$GeTe$_2$ with the rhombohedral space group ($R\bar{3}m$) shows the highest $T_c$, because the extra Fe layer in Fe$_{5-x}$GeTe$_2$ can considerably enhance the magnetic interaction[13]. Interestingly, the $T_c$ of Fe$_{5-x}$GeTe$_2$ is tunable by controlling the Fe deficiency content $x$, which ranges from 270 K to 363 K, demonstrating the pivotal role of Fe concentration in the magnetic exchange[14-16]. Additionally, the crucial role of Fe site could also be manifested by the helical magnetic ordering due to the noncentrosymmetry derived from the $\sqrt{3} \times \sqrt{3}$ ordering of the Fe(1)-Ge pair caused by the competition between the antisymmetric exchange interaction and the Heisenberg interaction[13]. To interpret the itinerant ferromagnetism, a classical Heisenberg model with Ruderman–Kittel–Kasuya–Yosida exchange is generally considered, which is expressed as:

$$H = \sum_{i,j} J_{ij} \vec{S}_i \cdot \vec{S}_j + \sum_i A(S_i^z)^2 \qquad (1)$$

where $\vec{S}_i$ is the spin operator on site $i$, $J_{ij}$ is the exchange coupling between spins on sites $i$ and $j$, and the parameter $A$ represents the single-ion perpendicular magnetocrystalline anisotropy arising from SOC.

In a system with reduced dimensionality, the SOC could be considerably enhanced[17]. As we mentioned above, SOC is conceived as a crucial role in stabilizing the perpendicular magnetic anisotropy down to one- or few-layer in 2D magnets, seen from eq.(1), which could also lift the chiral degeneracy, leading to the formation of topological magnetic configurations such as skyrmions through the Dzyaloshinskii-Moriya interaction [18]. As a quantum correction to the classical conductivity, the WAL effect can originate from strong SOC in the bulk materials and spin-momentum locking in the topological surface states of topological phases[19, 20], such as Bi-Se-Te-Sb system topological insulators (TIs) thin films[21-25] and 2D transitional metal chalcogenides (TMDs)[26, 27]. However, there have only very few reports of the WAL effect in magnets, and the observation in 2D vdW ferromagnets remains yet absent within our best knowledge. If the model described in eq.(1) is applicable for $Fe_{5-x}GeTe_2$, the SOC could play an important role. However, limited by the much less study on $Fe_{5-x}GeTe_2$ as compared with $Fe_3GeTe_2$, the underlying mechanism for the room temperature $T_c$ remains mysterious and the exact role of SOC has not been investigated yet.

In this work, we performed systematic magnetotransport measurements on the few-layer $Fe_{5-x}GeTe_2$ crystal, which unveiled a temperature dependent WAL effect caused by strong SOC. The WAL effect could persist up to a remarkably high temperature of about 120 K, which is the highest for such phenomena within our best knowledge. The magnetoconductance data were analyzed in details to deduce the spin-orbit scattering length $l_{SO}$ and the phase coherence length $l_{\varphi}$. Besides, we also observed a crossover from a tip to dip behavior in both magnetoresistance (MR) and magnetoconductance around 60 K, which is ascribed to the temperature induced Lifshitz transition that could induce sign change of carriers. The WAL effect disappears around 120 K likely due to a spin reorientation behavior.

**RESULTS AND DISCUSSION**

**Crystal structure, electrical and magnetotransport properties of $Fe_{5-x}GeTe_2$**. Figures 1(a) and (b) display the schematic crystal structure of $Fe_{5-x}GeTe_2$ viewed from the *b* and *c* axes, respectively. The structure consists of eight atom thick monolayers separated by Te atoms among each unit cell. The relatively weak vdW binding force among Te layers facilitates the exfoliation of the crystal into various layers. Seen in Figure 1(a), there are three occupations for Fe in the lattice, namely, Fe(1), Fe(2) and Fe(3), which are clearly labeled in the figure. The light purple circles labeled Fe(1) represent two possible positions of Fe(1) atoms, above or below a given Ge atom, corresponding to two possible Fe(1)-Ge split sites[13]. Figure 1(c) shows an optical microscope image of the typical device structure with a 20 $\mu$m scale bar, in which the thickness of $Fe_{5-x}GeTe_2$ crystal is about 29 nm as confirmed by the atomic force microscope measurements, i.e. which contains 10 layers since the thickness of a single layer is 2.9 nm. The resistivity of $Fe_{5-x}GeTe_2$ was measured in a standard four-probe configuration. Figure 1(d) shows the longitudinal resistance ($R_{xx}$) versus temperature ($T$) for the $Fe_{5-x}GeTe_2$ device shown in Figure 1(c) at the magnetic field ($B$) of 0 T, 1 T and 14 T, respectively, which show typical metallic behavior over the measured temperature range. Figure 1(e) shows the corresponding first derivation of $R_{xx}(T)$ curves. We have to note that the suggested charge order around 120 K is not visible except for some noise signals around this temperature, which might influence the resistivity negligibly [28].

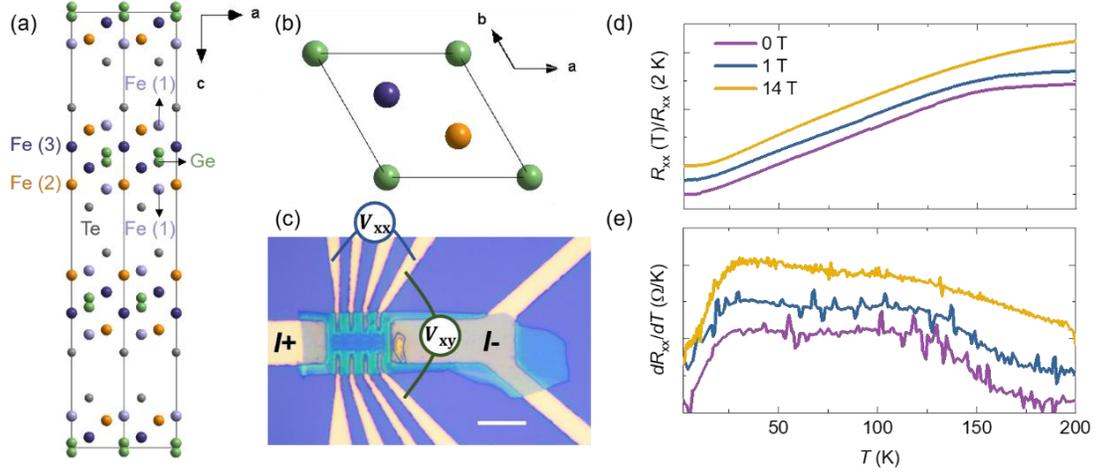

**Figure 1**. Schematic crystal structure of $Fe_{5-x}GeTe_2$ viewed from the **(a)** *b* axis and **(b)** *c* axis, respectively. **(c)** Optical image of a typical device with a Hall bar configuration. **(d)** The $R_{xx}(T)$ under 0 T, 1 T and 14 T magnetic field and **(e)** the corresponding first derivation of longitudinal resistance versus temperature at 0 T, 1 T and 14 T, respectively. All curves are offset to guide the eyes.

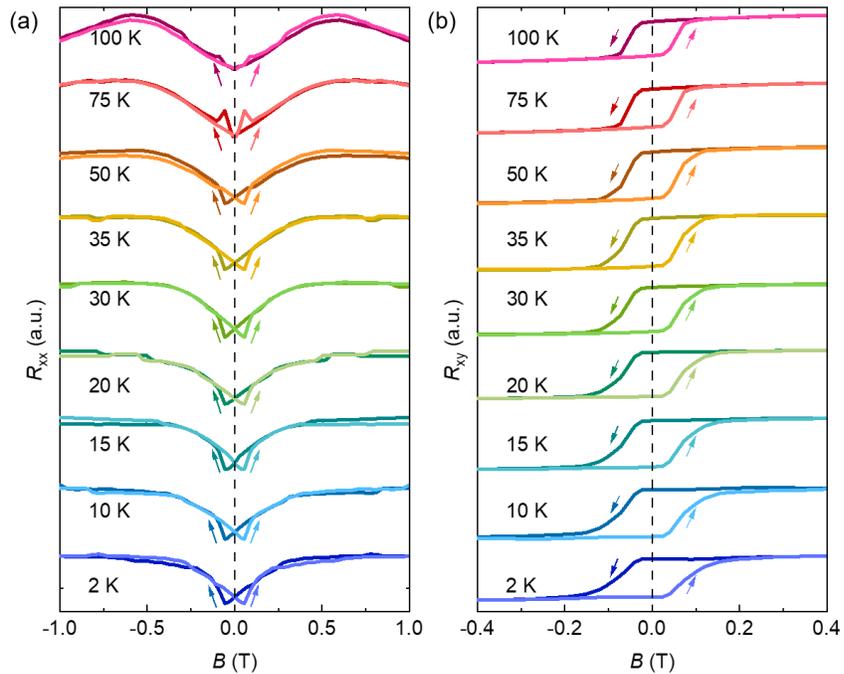

**Figure 2**. Temperature dependence of **(a)** longitudinal and **(b)** Hall sheet resistance at various magnetic fields.

Figure 2 shows the longitudinal and Hall sheet resistance for $Fe_{5-x}GeTe_2$ as a function of magnetic field within the temperature range of 2 - 100 K. The magnetic field was applied perpendicular to the current. The directions of the arrows in the figures denote the directions of the sweeping field. As shown in Figure 2(a), within the magnetic field range of $|B| < 0.1$ T, a hysteresis behavior in the MR is displayed, which might be caused by the gradual alignment of the magnetic domains by external magnetic field[29]. Strikingly, zero field dips signifying the WAL effect in the MR are displayed. With the increase of temperature, the MR dip at low magnetic field becomes broader and eventually disappears at 120 K with the gradually shortened phase coherent length. Seen in Figure 2(a), the MR shows a crossover from peak to dip between 50 and 75 K, likely indicative of a sign change of the carriers, which could be manifested by the angle-resolved photoemission spectroscopy (ARPES) measurements result as discussed later. Figure 2(b) presents the magnetic field dependence of Hall resistance ($R_{xy}$) at various temperatures corresponding to those in Figure 2(a). The nonlinear $R_{xy}(B)$ with clear hysteresis points to anomalous Hall effect in the few-layer $Fe_{5-x}GeTe_2$. With the increase of temperature, the coercive field $H_c$ and the saturation magnetic field $H_m$ gradually decrease, which will be discussed in more details later.

**Electronic structure of $Fe_{5-x}GeTe_2$**. To see more details about the electronic band structure of $Fe_{5-x}GeTe_2$, especially the crossover of peak to dip behavior in the MR between 50 and 75 K, the ARPES measurement results are presented in Figure 3. Figure 3(a) shows the constant energy contours at the binding energy of 0.05 eV under the $\sqrt{3} \times \sqrt{3}$ reconstruction [13]. The red and black solid lines represent the original and the reconstructed Brilloiun zones (BZs), respectively.

Since $Fe_{5-x}GeTe_2$ belongs to the space group $R\bar{3}m$, its two adjacent K points are not equivalent, which are distinguished by using $\bar{K}$ and $\bar{K}'$. The high symmetry points $\bar{\Gamma}$ and $\bar{M}$ are also marked in Figure 3(a). The band folding between $\Gamma$ and $\bar{K}$ points is clearly seen. From the dispersions along the high symmetry $\bar{K}$ - $\Gamma$ - $\bar{K}'$ directions, several hole-like bands could be identified. We focused on one of the hole bands, marked as band $\alpha$, to see its temperature evolution, which are shown in Figures 3(b)-3(c). We used a linear function to fit band $\alpha$ at different temperatures and used the slope of the linear function to calculate the Fermi velocity ($V_F$), which is plotted as a function of temperature in Figure 3(d). It is clear that a sharp jump occurs around 60 K, indicating a transition in the electronic band structure (especially the α band). The change of the other electronic bands near the Fermi level $E_F$ is more subtle to be analyzed. Although the origin of such transition remains unclear, the decrease of the (hole type) α band velocity is consistent with the change of the dominating carrier type from hole-like to electron-like. The result is in good agreement with the magnetotransport measurements showing the crossover from tip to dip behavior of the WAL effect. It should be notable that Lifshitz transition is commonly driven by chemical doing, external magnetic field or pressure, which whereas is scarcely induced by temperature. Temperature induced Lifshitz transition was recently observed in a series of topological semimetals such as $ZrTe_5$[30], $WTe_2$[31], $HfTe_5$[32], $ZrSiSe$[33], etc., due to the peculiar topological electronic band structures. The observation of temperature driven Lifshitz transition in $Fe_{5-x}GeTe_2$ would be very instructive for further understanding the complex charge order in 2D magnets.

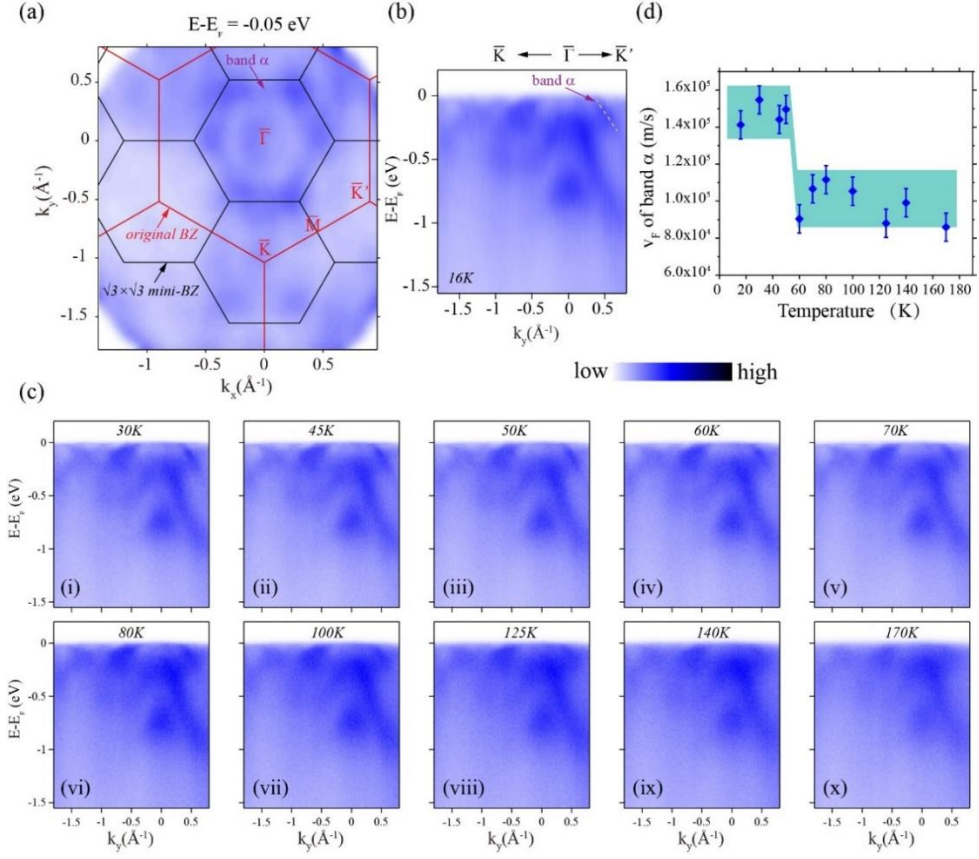

Figure 3. **(a)** Photoemission intensity map of constant energy contours in the $k_x$-$k_y$ plane at 0.05 eV below $E_F$. **(b)** and **(c)** Band dispersions along high-symmetry $\bar{K}$ - $\Gamma$ - $\bar{K}'$ direction with different temperatures. **(d)** Extracted $V_F$ of band $\alpha$ labeled in **(b)**, where the error bar corresponds to the energy resolution. The data were collected using photons with $h\nu$ = 112 eV.

These ARPES measured electronic structure is further supported by the first principles calculations which are shown in Figure 4(a). The calculated several hole-like bands around the Fermi level are consistent with the experimental ones with the intense state being about - 0.75 eV below. Since Fe(1) have two possible positions, the surface Fe atoms are either Fe(1) or Fe(5), as shown in Figure 4(b). According to the orbital projection of band structure presented in Figure 4(c), these hole-like bands are mainly derived from the hybrid states of surface Fe(1), Fe(5) and Te atoms, which may be related to the strong itinerant of Fe(1) and Fe(5) atoms to screen the inner other Fe atoms. The intense state which is about - 0.75 eV below the hole-like band is mainly originated from the *d* state of Fe(5) atom. In addition, the electronic structures of in-plane spin orientation and out-of-plane spin orientations are obviously different around $E_F$ for Fe(1)down-Fe(1)up-Fe(1)up order of Fe(1) layers (Figure S5), suggesting that the Lifshitz transition corresponds to the spin reorientation with temperature. The Fermi velocity across the Fermi energy around M' in the $\sqrt{3} \times \sqrt{3}$ mini Brillouin zone was calculated. The Fermi velocity of the out-of-plane spin orientation is $1.1 \times 10^5$ m/s, while the in-plane spin orientation (along *y* axis) is decreased to $0.8 \times 10^5$ m/s, which is similar to the sharp jump in experiment. The spin orientation is along the easy magnetized *z* axis at low temperature, and tends to the in-plane spin orientation or helical magnetic with the competition among magnetic anisotropy, Dzyaloshinskii-Moriya interaction and charge order as the temperature increases.

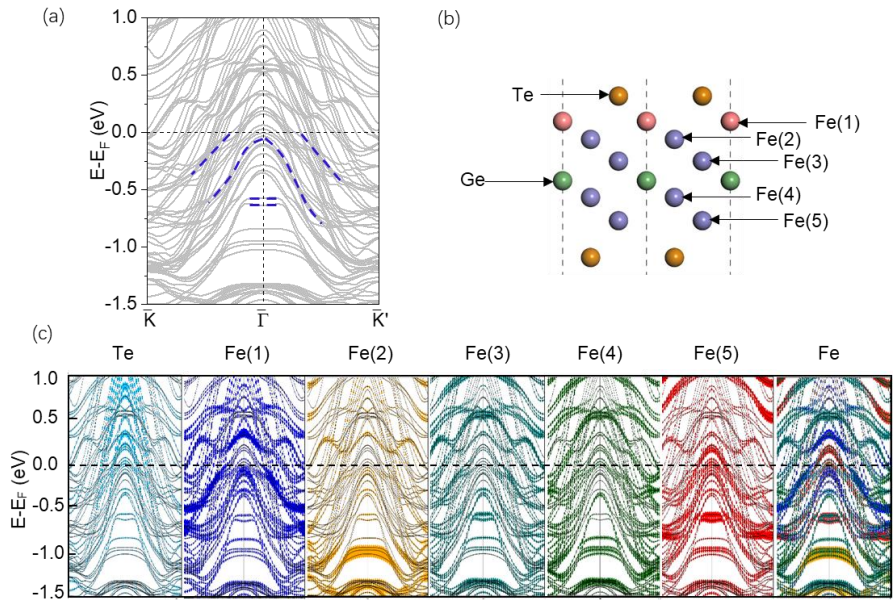

**Figure 4. (a)** Band structure along high-symmetry $\bar{K}$ - $\bar{\Gamma}$ - $\bar{K}'$ direction in original Brillouin zone of bulk. **(c)** Orbital-resolved band structures of Te atom and different Fe atoms marked in **(b)** along high-symmetry K - $\bar{\Gamma}$ - $\bar{K}'$ direction.

**Magnetoconductance of $Fe_{5-x}GeTe_2$**. To further understand the observed WAL effect in $Fe_{5-x}GeTe_2$, we conducted detailed analysis on the measured magnetoconductance. Figure 5(a) depicts the normalized sheet magnetoconductance $\Delta G_S = G_{xx}(B) - G_{xx}(0)$ from 2 K to 125 K as a function of perpendicular external magnetic field applied. In these plots, $\Delta G_S$ measured in unit of $e^2/h$, where $h$ is the Planck constant and $e$ is unit charge. WAL effect is suppressed by a perpendicular magnetic field, resulting in a negative magnetoconductance at least between 2 K and 10 K. With the increasing temperature, the localization behavior of $Fe_{5-x}GeTe_2$ is gradually altered. It is also clear that all magnetoconductance curves from 2 K to 75 K display sharp negative cusps around the zero magnetic field as characteristics of the WAL effect. With the increase of temperature, the cusps of $Fe_{5-x}GeTe_2$ are gradually suppressed and are not as sharp as that measured at low temperature. The width of the WAL cusps starts to increase and the amplitude decreases obviously as temperature rises from 50 K, which characterizes the SOC and phase coherent transport, respectively[34]. The WAL effect almost disappears as temperature is warmed up to ~ 120 K at which the magnetoconductance completely shows positive behavior. The WAL effect in $Fe_{5-x}GeTe_2$ is more pronounced in the regime of low temperature and low magnetic field, resembling the case of other materials.

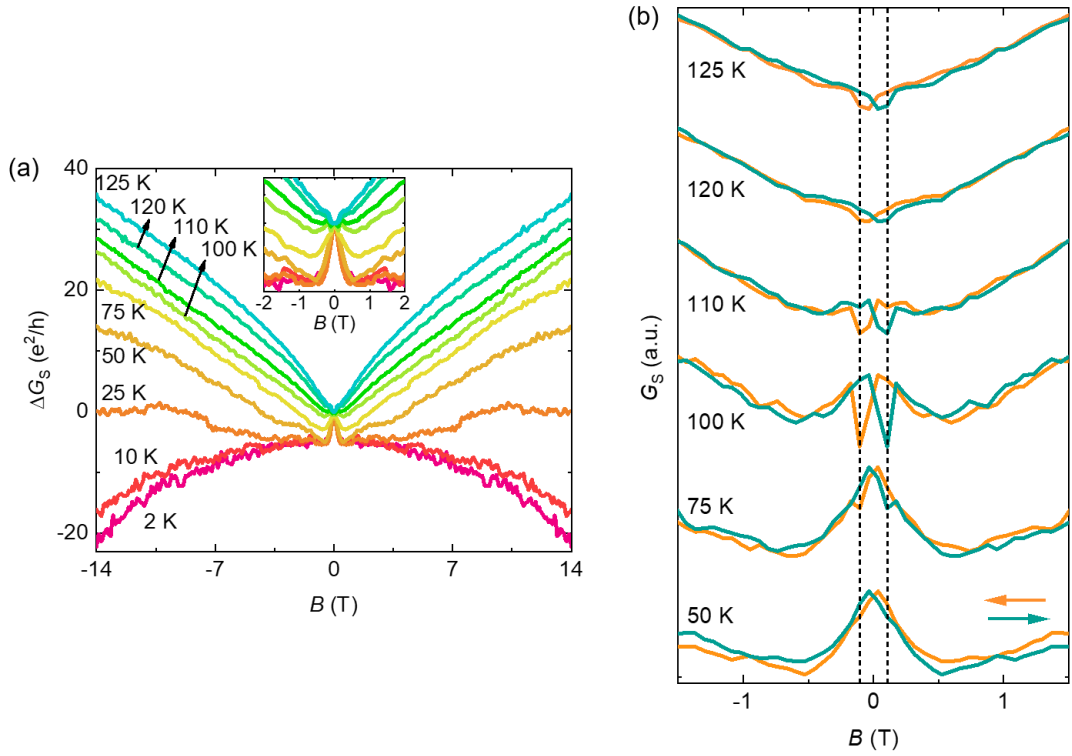

**Figure 5**. **(a)** Representative magnetoconductance curves exhibiting the WAL effect at different temperatures, which show that the magnitude of zero field cusps decreases gradually with the increasing temperature, and ultimately vanishes at around 120 K. Inset: Pronounced WAL cusps observed within the temperature range of 2 K to 125 K. **(b)** Details of the crossover from tips to dips behavior of the magnetoconductance. The arrows denote the directions of the sweeping magnetic field.

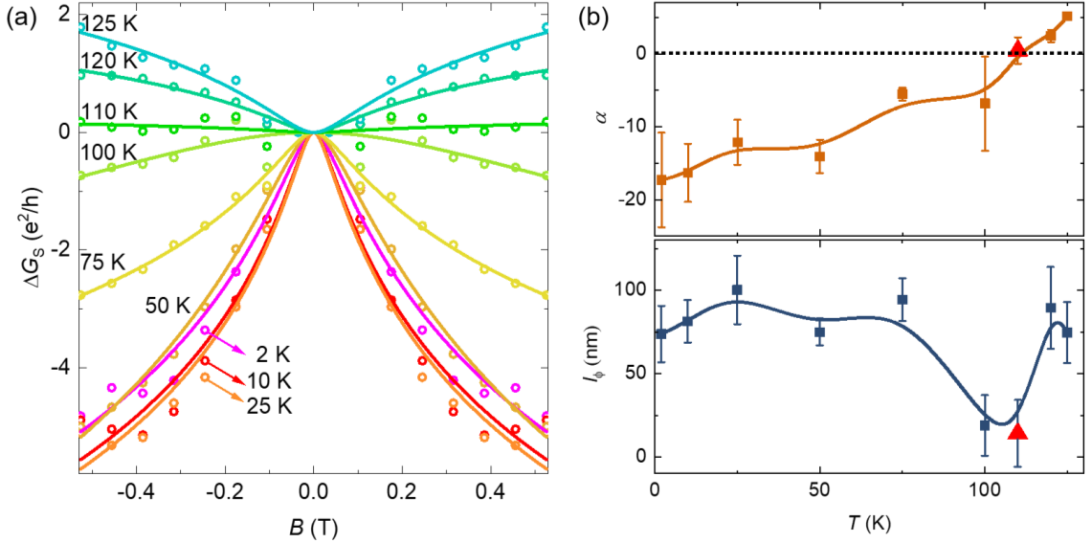

**Figure 6**. **(a)** Results of HLN fitting (solid lines) to the low field magnetoconductance (open circles) of $Fe_{5-x}GeTe_2$ in the temperature range of from 2 K to 125 K. **(b)** Evolution of coefficient $α$ and phase coherence length $l_φ$ with temperature. Vertical error bars represent the standard deviations of $α$ and $l_φ$ values.

To achieve more in-depth insights into the WAL effect in $Fe_{5-x}GeTe_2$, more quantitative analysis was carried out. Generally, the localization effect of 2D magnetoconductance can be described by the 2D Hikami-Larkin-Nagaoka (HLN) theory[19]. In the regime with low mobility and strong SOC, the variation of weak field conductance is expressed as:

$$\Delta G(H) = \frac{\alpha e^2}{2\pi^2 \hbar}\left[\Psi\left(\frac{1}{2} + \frac{\hbar}{4eBl_\varphi^2}\right) - ln\left(\frac{\hbar}{4eBl_\varphi^2}\right)\right] \quad (2)$$

where $\hbar$, $\Psi$, and $l_φ$ are the reduced plank constant, digamma function, and phase coherence length, respectively. The values of coefficient $α$ should be 1, 0, and -1/2 for the orthogonal, unitary, and symplectic cases, respectively. For TIs, $α$ denotes the number of conduction channels, which is theoretically predicted to be -0.5 for one conduction channel and -1 for a TI because the surface state exists on its both sides. Figure 6(a) displays the magnetoconductance at temperatures ranging from 2 K to 125 K (open circles) in the low field range from - 0.5 T to 0.5 T, which can be fitted fairly well by the 2D HLN model (solid lines), despite of the somewhat poor fit quality at high temperature. The HLN fitting exponents are summarized in Table S2. The fitting also gives the temperature dependence of phase coherence lengths $l_\varphi$ and coefficients $\alpha$, as summarized in Figure 6(b). It should be noted that the values of $\alpha$ are much larger than the theoretically predicted 2D values, which in fact have no any physical meanings, hinting that the the strong SOC induced WAL effect in $Fe_{5-x}GeTe_2$ is also contributed from of other 3D bulk channels. Such case were also found in TI $Bi_2Se_3$ crystals[35], $In_xSn_{1-x}Te$ nanoplates[36], TI $(Bi_{0.57}Sb_{0.43})_2Te_3$ thin films[37], half-Heusler semiconductor ScPdBi[38], and topological semimetal LuPdBi[39] *etc*. These results styrongly indicate that the origin of the WAL effect in $Fe_{5-x}GeTe_2$ is resulted from a combination of the 2D channels and 3D bulk states. Besides, in the upper panel of Figure 6(b), the sign of $\alpha$ changes with increasing temperature, likely indicating varied constitutions of conduction channels. The sign of $\alpha$ changes from negative to positive around 120 K, suggesting a smooth crossover from the sharp WAL behavior to conventional positive magnetoconductance behavior with increasing temperature[40]. In a system with anomalous Hall effect, the Hall coefficient usually contains two parts which are the ordinary Hall ($R_0$) and anomalous Hall ($\rho_{xy}^{AH}$ or $R_S$) resistivity. The $\rho_{xy}$ data are fitted with equation for $\rho_{xy}$, which is expressed as: $\rho_{xy} = R_0 H + \rho_{xy}^{AH}$, as illustrated in Figure S3. The roughly similar tendency of $\rho_{xy}^{AH}$ as $\alpha$ reveals the close relationship between the magnetization and WAL effect, which was also observed in Cr doped TI $Bi_2Se_3$ ultrathin films[40]. As shown in the lower panel of Figure 6(b), the temperature dependence of $l_\varphi$ indicates the dephasing process due to the enhancement of inelastic scattering with the increase of temperature.

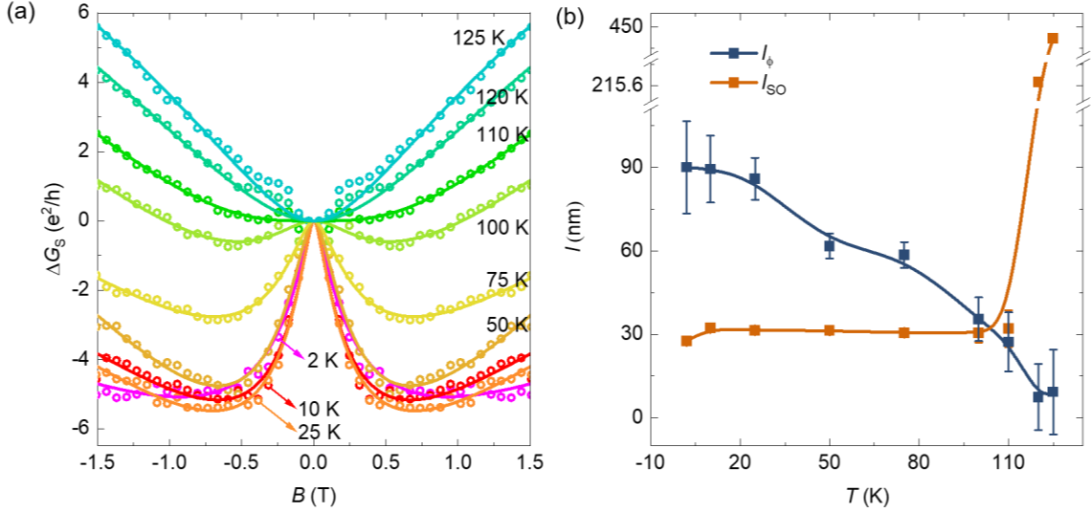

**Figure 7**. **(a)** Fitting curves (solid lines) of magnetoconductance (open circles) with the M-F model at various temperatures. **(b)** Extracted temperature dependent electron-phonon scattering length $l_\varphi$ and spin-orbit scattering length $l_{SO}$. Vertical error bars represent the standard deviation of $l_\varphi$ and $l_{SO}$ values.

On the other hand, for a 2D disordered electronic system, the field dependence of magnetoconductance could be calculated by using the Maekawa-Fukuyama (M-F) model[41] expressed as:

$$\Delta\sigma(T,B) = \frac{Ne^2}{\pi h}\left\{F\left(\frac{B}{B_\varphi+B_{SO}}\right) - \frac{1}{2}\left[F\left(\frac{B}{B_\varphi}\right) - F\left(\frac{B}{B_\varphi+2B_{SO}}\right)\right]\right\} \tag{3}$$

where $F(z) = \Psi\left(\frac{1}{2}+\frac{1}{z}\right) + ln(z)$ and $\Psi$ is the digamma function, $N$ is the number of independent conduction channels, $B_\varphi = \hbar/4el_\varphi^2$ and $B_{SO} = \hbar/4el_{SO}^2$, with $l_\varphi$ and $l_{SO}$ denoting the inelastic scattering induced dephasing length and spin-orbit scattering length. The results of fitting (solid lines) to the experimental data in the field range of ± 1.5 T by using this model are presented in Figure 7(a), in which the magnetoconductance is taken from 2 K to 125 K. The fitted parameters are summarized in Table S3. The fitting is basically satisfactory except the too large $N$ values which have no any physical meanings, similar as the situation of $\alpha$. The fact might also indicate that there are several conduction channels entangled with the 3D bulk states. Figure 7(b) shows the temperature dependence of extracted parameters $l_\varphi$ and $l_{SO}$. Below the temperature of around 120 K, $l_\varphi$ is larger than $l_{SO}$, implying that the spin-orbit scattering is dominated, which leads to the negative magnetoconductance and WAL phenomena. However, when the temperature is higher, $l_\varphi$ becomes smaller than $l_{SO}$ and the magnetoconductance and WAL effect becomes positive.

**Reflective magnetic circular dichroism (RMCD) measurements of $Fe_{5-x}GeTe_2$.** The RMCD measurements were performed on several few-layer $Fe_{5-x}GeTe_2$ crystals with varied thickness. As shown in Figure S4, with the reduction in thickness, the coercivity and ferromagnetism of $Fe_{5-x}GeTe_2$ are gradually enhanced. Figure 8 shows the results of RMCD measurement on two positions of $Fe_{5-x}GeTe_2$ crystals at different temperatures. The optical images of the few-layer $Fe_{5-x}GeTe_2$ crystals are presented in Figures 8(a) and 8(b) by the insets and the positions for RMCD measurements are marked by the green circles. It can be seen that the hysteresis and coercivity at 90 K and 120 K are basically similar, while those at 150 K changes significantly. Since the WAL effect disappears around 120 K, the spin reorientation induced competition between helimagnetism and other collinear properties in $Fe_{5-x}GeTe_2$ crystal might be the origin, which should be studied further in the future[13].

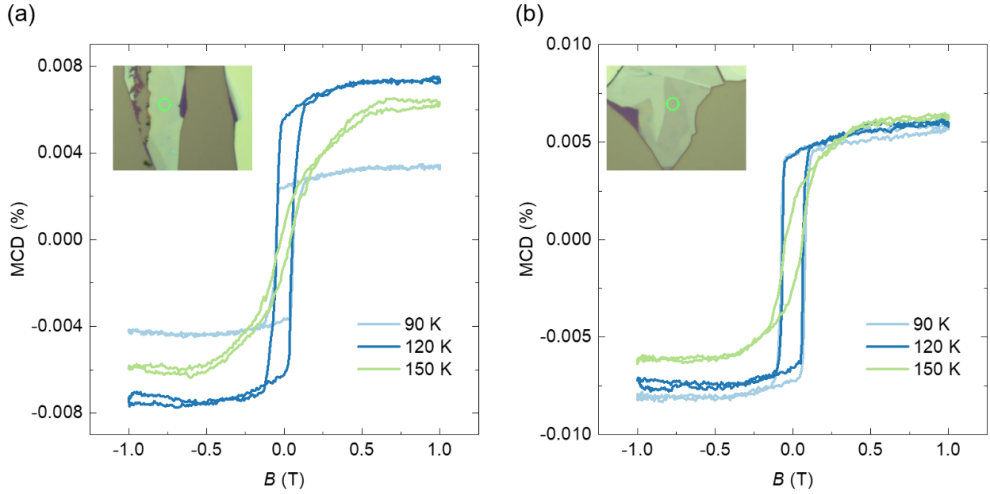

**Figure 8**. Results of temperature dependent RMCD for the few-layer $Fe_{5-x}GeTe_2$ crystals. Figures **(a)** and **(b)** correspond to the results measured on two different positions of the crystal marked by the green circles shown in the insets to each figure.

**Conclusions**

We observed a WAL effect in few-layer $Fe_{5-x}GeTe_2$ crystal due to strong SOC, which can even persist up to 120 K. By fitting the experimental data with the HLN and M-F models, the WAL effect in $Fe_{5-x}GeTe_2$ single crystals is ascribed to a combination of 2D channels and 3D bulk states. Additionally, we further studied the WAL behavior of $Fe_{5-x}GeTe_2$ in more details through ARPES and first principles calculations, which unveiled that the crossover of tip to dip behavior in both magnetoresistance and magnetoconductance could be attributed to a temperature induced Lifshitz transition. The RMCD measurements showed that the disappearance of WAL effect around 120 K is tightly related to the evolution of magnetism, which might due to a spin reorientation. Since the mechanism for WAL effect in magnetic system remains controversy and the WAL effect scarcely appeared in 2D magnets, our observation of the WAL effect persisted up to so high temperature in the vdW FM $Fe_{5-x}GeTe_2$ crystal provides an excellent platform for achieving more useful insights into this intriguing behavior. Additionally, the WAL effect in $Fe_{5-x}GeTe_2$ exposes the strong SOC effect, pointing to the possibility of the important role of SOC in the interaction Hamiltonian of such high $T_c$ itinerant ferromagnetism.

**Methods**

**Crystal growth and basic characterizations.** High-quality $Fe_{5-x}GeTe_2$ single crystals were grown by using the chemical vapor transport technique via employing iodine as the transport agent. Mixture of high purity elements including Fe, Ge, and Te with ratio of 6 : 1 : 2 were mixed and sealed in an evacuated quartz tube and slowly heated up to 700 ℃ in a tubular furnace. After stayed at the temperature for 7 days, the assembly was slowly cooled down to room temperature. The crystallographic phase and crystal quality were examined on a Bruker D8 single crystal X-ray diffractometer with Mo $K_α$ ($λ = 0.71073$ Å) at 300 K. The chemical compositions and uniformity of stoichiometry were checked by using the energy dispersive spectroscopy. More than 5 random areas on the surface of a crystal were measured, and the results are shown in Figure S1. The values were finally averaged, which are presented in Table S1, showing the compositions as Fe : Ge : Te = 4.6 : 1 : 2.1.

**Device fabrication and magnetotransport measurements.** The $Fe_{5-x}GeTe_2$ crystal was mechanically exfoliated into 10 layers in thickness. Standard UV photolithography with MJB4 aligner and lift-off techniques were used to fabricate the standard Hall bar configuration by deposition of 10/30 nm Ti/Au onto the 300 nm thick $SiO_2$/Si substrate. The crystal thickness was measured by atomic force microscope measurements. The few-layer $Fe_{5-x}GeTe_2$ single crystal was transferred using a silicone elastomer polydimethylsiloxane stamp to the prepared electrodes. All magnetotransport measurements were performed in a Quantum Design Physical Properties Measurement System, which has the lowest temperature of 1.8 K and the maximum magnetic field of 14 T. The standard Hall bar configuration was used in all samples to measure the longitudinal and Hall resistance. All the Hall and MR signals were normalized by subtracting the longitudinal and Hall resistance respectively over the positive and negative sweeping magnetic field range. The direction of magnetic field is perpendicular or parallel to the in-plane current of the devices.

**Density Functional Theory calculations.** DFT calculations were performed using the generalised gradient approximation for the exchange correlation potential with a plane-wave basis and the projector augmented wave method as implemented in the Vienna *ab initio* simulation package (VASP)[42-44]. The energy cut-off for the plane wave was set to 700 eV for structural relaxation and 500 eV for electronic structure calculations, and the *k*-point sampling of the first BZ is 15×15×2. The structures were fully relaxed until the residual force per atom was less than 0.001 eV/Å. In structural relaxation calculations, dispersion correction was considered in the optB86b functional for the exchange potential[45]. DFT+U method[46] was used to better describe the strong correlation system in structural relaxation and electronic structures. $U = 1$ eV is chosen in our calculations, and the lattice parameter *a* is 4.04 Å, *c* is 29.21 Å. In order to describe the two possible positions of Fe(1) in bulk, the distribution of Fe(1) in the three sublayer of the bulk phase is above, below and above Ge atom respectively.

**RMCD measurements.** RMCD signals were measured in a optical cryostat with temperature range from 1.5 K to 300 K and out-of-plane magnetic field up to 9 T. Polarized laser with the wavelength of 632.8 nm and the power of 15 ~ 20 $\mu$W was focused onto the flakes, which has the beam spot size around 0.7 $\mu$m. A photoelastic modulator was used to modulate the polarization of the incident light at the frequency of 50 kHz, and the signal was detected by a photodiode and analyzed by lock-in.

**ARPES measurements.** The ARPES measurements were performed at BL03U station at Shanghai Synchrotron Radiation Facility (SSRF), China. The $Fe_{5-x}GeTe_2$ single crystals were first elaborately selected and well prepared in air. Then they were cleaved in situ along the (001) surface without spoiling, and measured at a 15 K sample temperature in ultrahigh vacuum with a base pressure of more than $7 \times 10^{-11}$ Torr. Data were recorded by a VG Scienta DA30 analyzer. The angle resolution was 0.2° and the overall energy resolutions were better than 15 meV.

**AUTHOR INFORMATION**

**Corresponding Authors**

**Jun Li –**

*School of Physical Science and Technology, ShanghaiTech University, Shanghai 201210, China*

*ShanghaiTech Laboratory for Topological Physics, Shanghai 201210, China*

**Zhongkai Liu –**

*School of Physical Science and Technology, ShanghaiTech University, Shanghai 201210, China*

*ShanghaiTech Laboratory for Topological Physics, Shanghai 201210, China*

**Wei Ji –**

*Department of Physics and Beijing Key Laboratory of Opto-electronic Functional Materials & Micro-nano Devices, Renmin University of China, Beijing 100190, China*

**Yanfeng Guo –**

*School of Physical Science and Technology, ShanghaiTech University, Shanghai 201210, China*

**Authors**

**Zhengxian Li –**

*School of Physical Science and Technology, ShanghaiTech University, Shanghai 201210, China*

**Kui Huang –**

*School of Physical Science and Technology, ShanghaiTech University, Shanghai 201210, China*

**Deping Guo –**

*Department of Physics and Beijing Key Laboratory of Opto-electronic Functional Materials & Micro-nano Devices, Renmin University of China, Beijing 100190, China*

**Guodong Ma –**

*National Laboratory of Solid State Microstructures and School of Physics, Nanjing University, Nanjing 210093, China*

**Xiaolei Liu –**

*School of Physical Science and Technology, ShanghaiTech University, Shanghai 201210, China*

**Yueshen Wu –**

*School of Physical Science and Technology, ShanghaiTech University, Shanghai 201210, China*

**Jian Yuan –**

*School of Physical Science and Technology, ShanghaiTech University, Shanghai 201210, China*

**Zicheng Tao –**

*School of Physical Science and Technology, ShanghaiTech University, Shanghai 201210, China*

**Xia Wang –**

*School of Physical Science and Technology, ShanghaiTech University, Shanghai 201210, China*

*Analytical Instrumentation Center, School of Physical Science and Technology, ShanghaiTech University, Shanghai 201210, China*

**Zhiqiang Zou –**

*School of Physical Science and Technology, ShanghaiTech University, Shanghai 201210, China*

*Analytical Instrumentation Center, School of Physical Science and Technology, ShanghaiTech University, Shanghai 201210, China*

**Na Yu –**

*School of Physical Science and Technology, ShanghaiTech University, Shanghai 201210, China*

*Analytical Instrumentation Center, School of Physical Science and Technology, ShanghaiTech University, Shanghai 201210, China*

**Geliang Yu –**

*National Laboratory of Solid State Microstructures and School of Physics, Nanjing University, Nanjing 210093, China*

**Jimin Xue –**

*School of Physical Science and Technology, ShanghaiTech University, Shanghai 201210, China*

**Author Contributions**

Y.F.G. conceived the project. Z.X.L. grew the crystals, did the basic chracterizations, fabricated the devices and measured the magnetotransport properties with the help from Z.C.T., J.Y.,

X.L.L., X.W., Z.Q.Z., N.Y., B.B.W. and Y.S.W. directed by J.M.X., Y.F.G. and J.L. K.H. performed the ARPES measurements directed by Z.K.L. D.P.G. did the calculations directed by W.J. G.D.M. measured the RMCD directed by G.L.Y. All authors contributed to the writing of the manuscript.

**ACKNOWLEDGEMENTS**

This work was supported by the National Science Foundation of China (Grant Nos.92065201, 11874264). Y.F.G. acknowledges the starting grant of ShanghaiTech University and the open projects from State Key Laboratory of Surface Physics and Department of Physics, Fudan University (Grant No. KF2020_09) and National Laboratory of Solid State Microstructures, Nanjing University (grant No. M34015). W.J. acknowledges support by the National Key Research and Development Program of China (No. 2018YFE0202700), the National Natural Science Foundation of China (Nos. 11622437, 61674171 and 11974422), and the Strategic Priority Research Program of the Chinese Academy of Sciences (No. XDB30000000). B.W., J.X. and Z.K.L. acknowledge financial support from the Ministry of Science and Technology of China (2017YFA0305400) and the Strategic Priority Research Program of Chinese Academy of Sciences (XDA18010000). J.L. acknowledges the National Natural Science Foundation of China (Grants No. 61771234 and No. 12004251), the Natural Science Foundation of Shanghai (Grant No. 20ZR1436100), the Science and Technology Commission of Shanghai Municipality. The authors also thank the support from Analytical Instrumentation Center (#SPST-AIC10112914) and the Soft Matter Nanofab (SMN180827), SPST, ShanghaiTech University.

## ASSOCIATED CONTENT

**Supporting information Available**

The following files are available free of charge. Additional figures and data to support the results in the main text (PDF).

Results of the energy dispersive X-ray spectrum (EDS) measurements; Angle dependent magnetoresistance of bulk $Fe_{5-x}GeTe_2$ crystal; Anomalous Hall $\rho_{xy}^{AH}$ contribution to Hall resistivity $\rho_{xy}$ as a function of temperature; RMCD measurement results of $Fe_{5-x}GeTe_2$ samples with different thickness at 1.6 K; Band structure along high-symmetry M-Γ-M' direction in $\sqrt{3} \times \sqrt{3}$ mini Brillouin zone of monolayer; The HNL fitting exponents; The M-F fitting exponents.

## REFERENCES

(1) Novoselov, K. S.; Geim, A. K.; Morozov, S. V.; Jiang, D.-e.; Zhang, Y.; Dubonos, S. V.; Grigorieva, I. V.; Firsov, A. A. Electric field effect in atomically thin carbon films. *Science* **2004,** 306, (5696), 666-669.

(2) Zhang, Y.; Tan, Y.-W.; Stormer, H. L.; Kim, P. Experimental observation of the quantum Hall effect and Berry's phase in graphene. *Nature* **2005,** 438, (7065), 201-204.

(3) Mak, K. F.; Lee, C.; Hone, J.; Shan, J.; Heinz, T. F. Atomically thin $MoS_2$: a new direct-gap semiconductor. *Phys. Rev. Lett.* **2010,** 105, (13), 136805.

(4) Lu, J.; Zheliuk, O.; Leermakers, I.; Yuan, N. F.; Zeitler, U.; Law, K. T.; Ye, J. Evidence for two-dimensional Ising superconductivity in gated $MoS_2$. *Science* **2015,** 350, (6266), 1353-1357.

(5) Xi, X.; Zhao, L.; Wang, Z.; Berger, H.; Forró, L.; Shan, J.; Mak, K. F. Strongly enhanced charge-density-wave order in monolayer NbSe$_2$. *Nat. Nanotech.* **2015,** 10, (9), 765-769.

(6) Li, L.; O'Farrell, E.; Loh, K.; Eda, G.; Özyilmaz, B.; Neto, A. C. Controlling many-body states by the electric-field effect in a two-dimensional material. *Nature* **2016,** 529, (7585), 185-189.

(7) Mermin, N. D.; Wagner, H. Absence of ferromagnetism or antiferromagnetism in one-or two-dimensional isotropic Heisenberg models. *Phys. Rev. Lett.* **1966,** 17, (22), 1133.

(8) Huang, B.; Clark, G.; Navarro-Moratalla, E.; Klein, D. R.; Cheng, R.; Seyler, K. L.; Zhong, D.; Schmidgall, E.; McGuire, M. A.; Cobden, D. H.; Yao, W.; Xiao, D.; Jarillo-Herrero, P.; Xu, X. Layer-dependent ferromagnetism in a van der Waals crystal down to the monolayer limit. *Nature* **2017,** 546, (7657), 270-273.

(9) Gong, C.; Li, L.; Li, Z.; Ji, H.; Stern, A.; Xia, Y.; Cao, T.; Bao, W.; Wang, C.; Wang, Y. Discovery of intrinsic ferromagnetism in two-dimensional van der Waals crystals. *Nature* **2017,** 546, (7657), 265-269.

(10) Lin, M.-W.; Zhuang, H. L.; Yan, J.; Ward, T. Z.; Puretzky, A. A.; Rouleau, C. M.; Gai, Z.; Liang, L.; Meunier, V.; Sumpter, B. G. Ultrathin nanosheets of CrSiTe$_3$: a semiconducting two-dimensional ferromagnetic material. *J. Mater. Chem. C* **2016,** 4, (2), 315-322.

(11) Bonilla, M.; Kolekar, S.; Ma, Y.; Diaz, H. C.; Kalappattil, V.; Das, R.; Eggers, T.; Gutierrez, H. R.; Phan, M.-H.; Batzill, M. Strong room-temperature ferromagnetism in VSe$_2$ monolayers on van der Waals substrates. *Nat. Nanotech.* **2018,** 13, (4), 289-293.

(12) O'Hara, D. J.; Zhu, T.; Trout, A. H.; Ahmed, A. S.; Luo, Y. K.; Lee, C. H.; Brenner, M. R.; Rajan, S.; Gupta, J. A.; McComb, D. W. Room temperature intrinsic ferromagnetism in epitaxial manganese selenide films in the monolayer limit. *Nano Lett.* **2018,** 18, (5), 3125-3131.

(13) Ly, T. T.; Park, J.; Kim, K.; Ahn, H.-B.; Lee, N. J.; Kim, K.; Park, T.-E.; Duvjir, G.; Lam, N. H.; Jang, K.; You, C.-Y.; Jo, Y.; Kim, S. K.; Lee, C.; Kim, S.; Kim, J. Direct Observation of Fe-Ge Ordering in $Fe_{5-x}GeTe_2$ Crystals and Resultant Helimagnetism. *Adv. Funct. Mater.* **2021,** 31, (17), 2009758.

(14) May, A. F.; Ovchinnikov, D.; Zheng, Q.; Hermann, R.; Calder, S.; Huang, B.; Fei, Z.; Liu, Y.; Xu, X.; McGuire, M. A. Ferromagnetism Near Room Temperature in the Cleavable van der Waals Crystal $Fe_5GeTe_2$. *ACS Nano* **2019,** 13, (4), 4436-4442.

(15) May, A. F.; Bridges, C. A.; McGuire, M. A. Physical properties and thermal stability of $Fe_{5-x}GeTe_2$ single crystals. *Phys. Rev. Mater.* **2019,** 3, (10), 104401.

(16) Tian, C.; Pan, F.; Xu, S.; Ai, K.; Xia, T.; Cheng, P. Tunable magnetic properties in van der Waals crystals $(Fe_{1-x}Co_x)_5GeTe_2$. *Appl. Phys. Lett.* **2020,** 116, (20), 202402.

(17) Soumyanarayanan, A.; Reyren, N.; Fert, A.; Panagopoulos, C. Emergent phenomena induced by spin-orbit coupling at surfaces and interfaces. *Nature* **2016,** 539, (7630), 509-517.

(18) Gayles, J.; Freimuth, F.; Schena, T.; Lani, G.; Mavropoulos, P.; Duine, R. A.; Blügel, S.; Sinova, J.; Mokrousov, Y. Dzyaloshinskii-Moriya Interaction and Hall Effects in the Skyrmion Phase of $Mn_{1-x}Fe_xGe$. *Phys. Rev. Lett.* **2015,** 115, (3), 036602.

(19) Hikami, S.; Larkin, A. I.; Nagaoka, Y. Spin-orbit interaction and magnetoresistance in the two dimensional random system. *Prog. Theor. Phys.* **1980,** 63, (2), 707-710.

(20) Nomura, K.; Koshino, M.; Ryu, S. Topological Delocalization of Two-Dimensional Massless Dirac Fermions. *Phys. Rev. Lett.* **2007,** 99, (14), 146806.

(21) Cha, J. J.; Kong, D.; Hong, S.-S.; Analytis, J. G.; Lai, K.; Cui, Y. Weak antilocalization in $Bi_2(Se_xTe_{1-x})_3$ nanoribbons and nanoplates. *Nano Lett.* **2012,** 12, (2), 1107-1111.

(22) Bao, L.; He, L.; Meyer, N.; Kou, X.; Zhang, P.; Chen, Z.-g.; Fedorov, A. V.; Zou, J.; Riedemann, T. M.; Lograsso, T. A. Weak anti-localization and quantum oscillations of surface states in topological insulator $Bi_2Se_2Te$. *Sci. Rep.* **2012,** 2, (1), 1-7.

(23) Chen, J.; Qin, H.; Yang, F.; Liu, J.; Guan, T.; Qu, F.; Zhang, G.; Shi, J.; Xie, X.; Yang, C. Gate-voltage control of chemical potential and weak antilocalization in $Bi_2Se_3$. *Phys. Rev. Lett.* **2010,** 105, (17), 176602.

(24) He, H.-T.; Wang, G.; Zhang, T.; Sou, I.-K.; Wong, G. K.; Wang, J.-N.; Lu, H.-Z.; Shen, S.-Q.; Zhang, F.-C. Impurity effect on weak antilocalization in the topological insulator $Bi_2Te_3$. *Phys. Rev. Lett.* **2011,** 106, (16), 166805.

(25) Li, H.; Wang, H.-W.; Li, Y.; Zhang, H.; Zhang, S.; Pan, X.-C.; Jia, B.; Song, F.; Wang, J. Quantitative Analysis of Weak Antilocalization Effect of Topological Surface States in Topological Insulator $BiSbTeSe_2$. *Nano Lett.* **2019,** 19, (4), 2450-2455.

(26) Liu, H.; Bao, L.; Zhou, Z.; Che, B.; Zhang, R.; Bian, C.; Ma, R.; Wu, L.; Yang, H.; Li, J.; Gu, C.; Shen, C. M.; Du, S.; Gao, H. J. Quasi-2D Transport and Weak Antilocalization Effect in Few-layered $VSe_2$. *Nano Lett.* **2019,** 19, (7), 4551-4559.

(27) Naylor, C. H.; Parkin, W. M.; Ping, J.; Gao, Z.; Zhou, Y. R.; Kim, Y.; Streller, F.; Carpick, R. W.; Rappe, A. M.; Drndic, M.; Kikkawa, J. M.; Johnson, A. T. Monolayer Single-Crystal 1T'-$MoTe_2$ Grown by Chemical Vapor Deposition Exhibits Weak Antilocalization Effect. *Nano Lett.* **2016,** 16, (7), 4297-304.

(28) Wu, X.; Lei, L.; Yin, Q.; Zhao, N.-N.; Li, M.; Wang, Z.; Liu, Q.; Song, W.; Ma, H.; Ding, P. Direct observation of competition between charge order and itinerant ferromagnetism in vdW crystal $Fe_{5-x}GeTe_2$. **2021**.

(29) Chiu, S.-P.; Yamanouchi, M.; Oyamada, T.; Ohta, H.; Lin, J.-J. Gate tunable spin-orbit coupling and weak antilocalization effect in an epitaxial La$_{2/3}$Sr$_{1/3}$MnO$_3$ thin film. *Phys. Rev. B* **2017,** 96, (8), 085143.

(30) Zhang, Y.; Wang, C.; Yu, L.; Liu, G.; Liang, A.; Huang, J.; Nie, S.; Sun, X.; Zhang, Y.; Shen, B.; Liu, J.; Weng, H.; Zhao, L.; Chen, G.; Jia, X.; Hu, C.; Ding, Y.; Zhao, W.; Gao, Q.; Li, C.; He, S.; Zhao, L.; Zhang, F.; Zhang, S.; Yang, F.; Wang, Z.; Peng, Q.; Dai, X.; Fang, Z.; Xu, Z.; Chen, C.; Zhou, X. J. Electronic evidence of temperature-induced Lifshitz transition and topological nature in ZrTe$_5$. *Nat. Commun.* **2017,** 8, (1), 15512.

(31) Wu, Y.; Jo, N. H.; Ochi, M.; Huang, L.; Mou, D.; Bud'ko, S. L.; Canfield, P. C.; Trivedi, N.; Arita, R.; Kaminski, A. Temperature-Induced Lifshitz Transition in WTe$_2$. *Phys. Rev. Lett.* **2015,** 115, (16), 166602.

(32) Zhang, Y.; Wang, C.; Liu, G.; Liang, A.; Zhao, L.; Huang, J.; Gao, Q.; Shen, B.; Liu, J.; Hu, C.; Zhao, W.; Chen, G.; Jia, X.; Yu, L.; Zhao, L.; He, S.; Zhang, F.; Zhang, S.; Yang, F.; Wang, Z.; Peng, Q.; Xu, Z.; Chen, C.; Zhou, X. Temperature-induced Lifshitz transition in topological insulator candidate HfTe$_5$. *Sci. Bull.* **2017,** 62, (13), 950-956.

(33) Chen, F. C.; Fei, Y.; Li, S. J.; Wang, Q.; Luo, X.; Yan, J.; Lu, W. J.; Tong, P.; Song, W. H.; Zhu, X. B.; Zhang, L.; Zhou, H. B.; Zheng, F. W.; Zhang, P.; Lichtenstein, A. L.; Katsnelson, M. I.; Yin, Y.; Hao, N.; Sun, Y. P. Temperature-Induced Lifshitz Transition and Possible Excitonic Instability in ZrSiSe. *Phys. Rev. Lett.* **2020,** 124, (23), 236601.

(34) Zhang, Y.; Xue, F.; Tang, C.; Li, J.; Liao, L.; Li, L.; Liu, X.; Yang, Y.; Song, C.; Kou, X. Highly Efficient Electric-Field Control of Giant Rashba Spin–Orbit Coupling in Lattice-Matched InSb/CdTe Heterostructures. *ACS Nano* **2020,** 14, (12), 17396-17404.

(35) Checkelsky, J. G.; Hor, Y. S.; Liu, M. H.; Qu, D. X.; Cava, R. J.; Ong, N. P. Quantum Interference in Macroscopic Crystals of Nonmetallic $Bi_2Se_3$. *Phys. Rev. Lett.* **2009,** 103, (24), 246601.

(36) Shen, J.; Xie, Y.; Cha, J. J. Revealing Surface States in In-Doped SnTe Nanoplates with Low Bulk Mobility. *Nano Lett.* **2015,** 15, (6), 3827-3832.

(37) Lang, M.; He, L.; Kou, X.; Upadhyaya, P.; Fan, Y.; Chu, H.; Jiang, Y.; Bardarson, J. H.; Jiang, W.; Choi, E. S.; Wang, Y.; Yeh, N. C.; Moore, J.; Wang, K. L. Competing weak localization and weak antilocalization in ultrathin topological insulators. *Nano Lett.* **2013,** 13, (1), 48-53.

(38) Zhang, J.; Hou, Z.; Zhang, C.; Chen, J.; Li, P.; Wen, Y.; Zhang, Q.; Wang, W.; Zhang, X. Weak antilocalization effect and high-pressure transport properties of ScPdBi single crystal. *Appl. Phys. Lett.* **2019,** 115, (17), 172407.

(39) Xu, G.; Wang, W.; Zhang, X.; Du, Y.; Liu, E.; Wang, S.; Wu, G.; Liu, Z.; Zhang, X. X. Weak antilocalization effect and noncentrosymmetric superconductivity in a topologically nontrivial semimetal LuPdBi. *Sci. Rep.* **2014,** 4, (1), 5709.

(40) Liu, M.; Zhang, J.; Chang, C. Z.; Zhang, Z.; Feng, X.; Li, K.; He, K.; Wang, L. L.; Chen, X.; Dai, X.; Fang, Z.; Xue, Q. K.; Ma, X.; Wang, Y. Crossover between weak antilocalization and weak localization in a magnetically doped topological insulator. *Phys. Rev. Lett.* **2012,** 108, (3), 036805.

(41) Maekawa, S.; Fukuyama, H. Magnetoresistance in two-dimensional disordered systems: effects of Zeeman splitting and spin-orbit scattering. *J. Phys. Soc. Jpn.* **1981,** 50, (8), 2516-2524.

(42) Blöchl, P. E. Projector augmented-wave method. *Phys. Rev. B* **1994,** 50, (24), 17953-17979.

(43)    Kresse, G.; Furthmüller, J. Efficient iterative schemes for ab initio total-energy calculations using a plane-wave basis set. *Phys. Rev. B* **1996,** 54, (16), 11169-11186.

(44)    Kresse, G.; Joubert, D. From ultrasoft pseudopotentials to the projector augmented-wave method. *Phys. Rev. B* **1999,** 59, (3), 1758-1775.

(45)    Klimeš, J.; Bowler, D. R.; Michaelides, A. Van der Waals density functionals applied to solids. *Phys. Rev. B* **2011,** 83, (19), 195131.

(46)    Anisimov, V. I.; Aryasetiawan, F.; Lichtenstein, A. I. First-principles calculations of the electronic structure and spectra of strongly correlated systems: the LDA+Umethod. *J. Phys.: Condens. Matter* **1997,** 9, (4), 767-808.

# Supporting Information

# Weak Antilocalization Effect up to ~120 K in the van der Waals Crystal $Fe_{5-x}GeTe_2$ with Near Room Temperature Ferromagnetism

Authors: *Zhengxian Li[1,†], Kui Huang[1,†], Deping Guo[2,†], Guodong Ma[3], Xiaolei Liu[1], Yueshen Wu[1], Jian Yuan[1], Zicheng Tao[1], Binbin Wang[1], Xia Wang[1,4], Zhiqiang Zou[1,4], Na Yu[1,4], Geliang Yu[3], Jiamin Xue[1], Jun Li[1,5]\*, Zhongkai Liu[1,5]\*, Wei Ji[2]\*, Yanfeng Guo[1,3,6]\*.*

AUTHOR ADDRESS

[1]School of Physical Science and Technology, ShanghaiTech University, Shanghai 201210, China

[2]Department of Physics and Beijing Key Laboratory of Opto-electronic Functional Materials & Micro-nano Devices, Renmin University of China, Beijing 100190, China

[3]National Laboratory of Solid State Microstructures and School of Physics, Nanjing University, Nanjing 210093, China

[4]Analytical Instrumentation Center, School of Physical Science and Technology, ShanghaiTech University, Shanghai 201210, China

[5]ShanghaiTech Laboratory for Topological Physics, Shanghai 201210, China

[6]State Key Laboratory of Surface Physics and Department of Physics, Fudan University, Shanghai 200433, China

[†]These authors contributed equally to this work.

*Email address: lijun3@shanghaitech.edu.cn, liuzhk@@shanghaitech.edu.cn, wji@ruc.edu.cn, guoyf@shanghaitech.edu.cn.

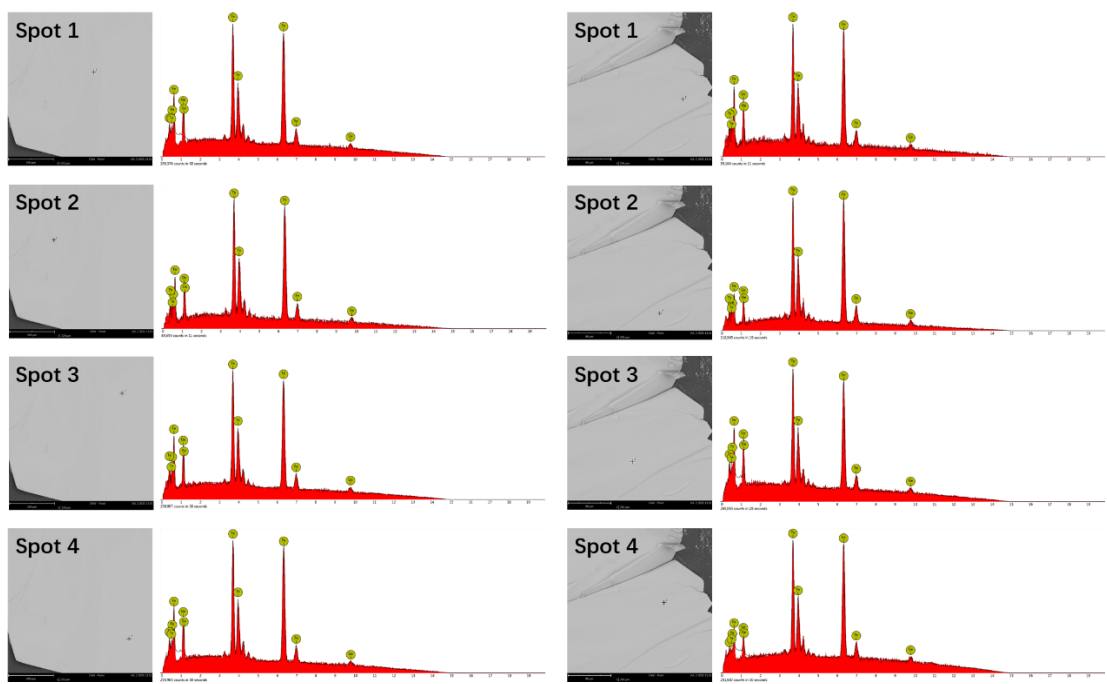

**Figure S1.** Results of the energy dispersive X-ray spectrum (EDS) measurements on different points of the $Fe_{5-x}GeTe_2$ crystal.

Table S1. The averaged values of the EDS results of the Fe$_{5-x}$GeTe$_2$ single crystal.

| Element Number | Element Symbol | Element Name | Atomic Conc. | Weight Conc. |
|---|---|---|---|---|
| 26 | Fe | Iron | 59.22 | 42.37 |
| 52 | Te | Tellurium | 27.97 | 45.70 |
| 32 | Ge | Germanium | 12.82 | 11.93 |

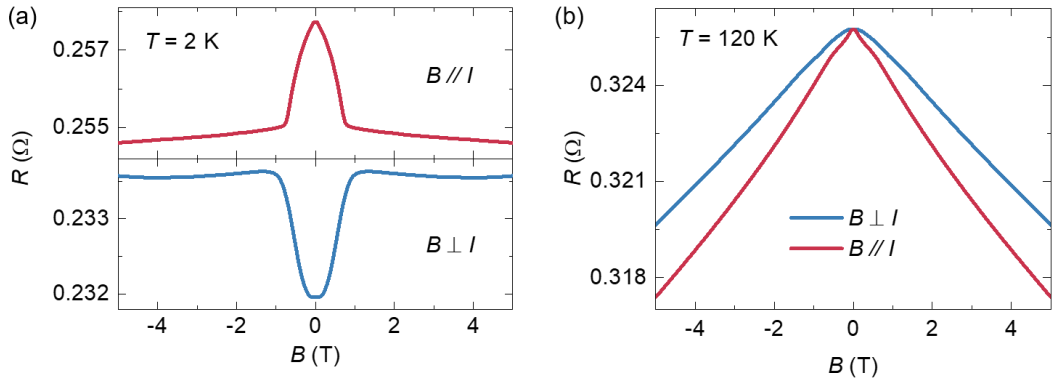

**Figure S2.** Angle dependent magnetoresistance of bulk Fe$_{5-x}$GeTe$_2$ crystal measured at (a) 2 K and (b) 120 K within the magnetic field range of $|B| < 5$ T. Magnetic fields were applied perpendicular and parallel to the current, respectively.

To further expose the origin of the WAL effect in few-layer Fe$_{5-x}$GeTe$_2$, angle dependence of MR measurements were performed by changing the direction of the applied magnetic field. Figure S2 presents the data measured at 2 K and 120 K, respectively. The angle dependent behavior seen in Figure S2 may strongly indicate a 2D nature of the electrical transport in $Fe_{5-x}GeTe_2$ because if the WAL effect is mainly caused by SOC in a 3D bulk channel, the MR should be independent of the tilt angles of the magnetic field [1-3]. It is noted that the angle dependent MR behaviors have also been observed in the 2D vdW magnetic material $CrTe_2$ [4].

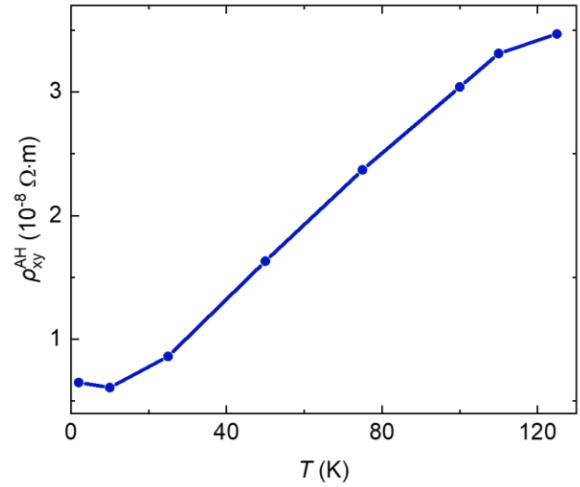

**Figure S3.** Anomalous Hall $\rho_{xy}^{AH}$ contribution to Hall resistivity $\rho_{xy}$ as a function of temperature.

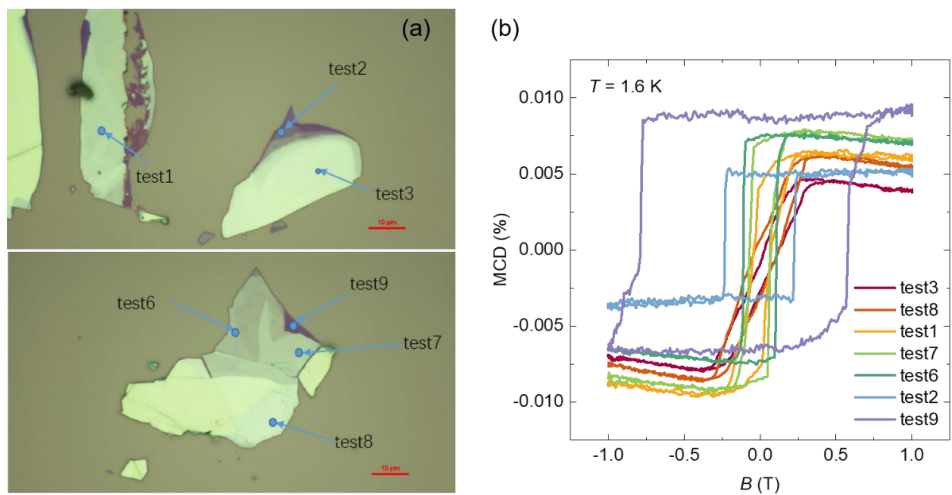

**Figure S4.** (a) Optical images of the few-layer $Fe_{5-x}GeTe_2$ crystals on the silicon wafer. (b) MCD measurement results of $Fe_{5-x}GeTe_2$ samples with different thickness at 1.6 K.

Figure S4(a) shows the optical images of the few-layer $Fe_{5-x}GeTe_2$ crystal on the silicon wafer. The measured points on the sample with varied thickness are labeled as 3, 8, 1, 7, 6, 2, 9, respectively, where the sample in deep purple is the thinnest and the ones in bright yellow are relatively thicker. All the MCD measurements performed at these points were at 1.6 K, as shown in Figure S4(b). It is obvious that as the thickness decreases, both the coercivity and the ferromagnetism of $Fe_{5-x}GeTe_2$ crystal are gradually weakened.

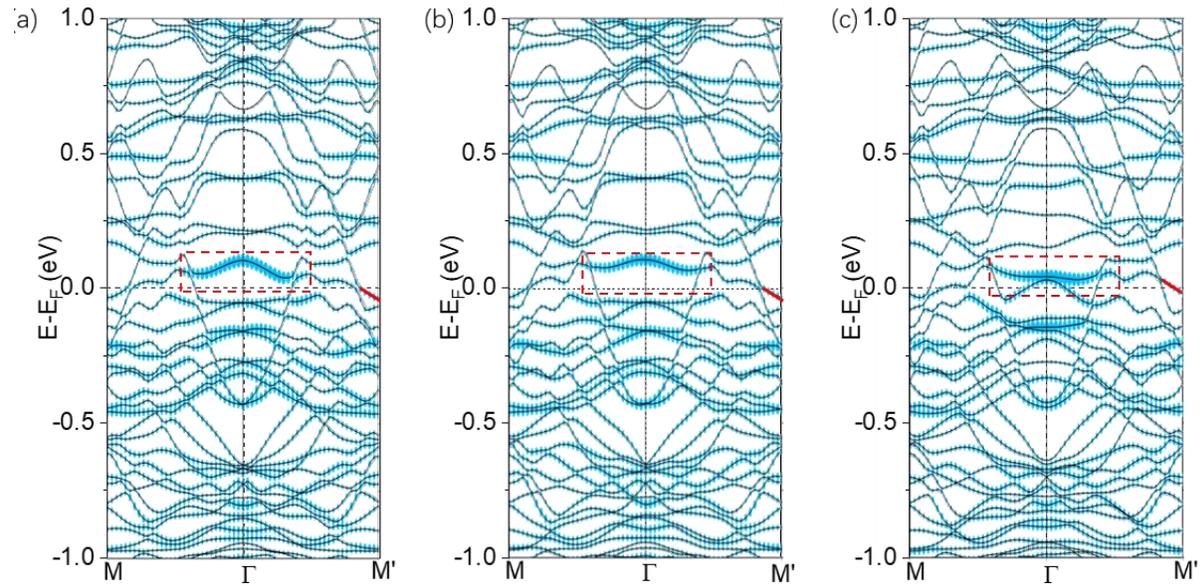

**Figure S5.** Band structure along high-symmetry M-Γ-M' direction in $\sqrt{3} \times \sqrt{3}$ mini Brillouin zone of monolayer. (a) spin orientation along *x* axis. (b) spin orientation along *y* axis. (c) spin orientation along *z* axis. The Fermi velocity of band marked red solid line is calculated. The red dotted box marks the obvious different energy bands about the Fermi level in different spin orientation

Table S2. HLN fitting exponents.

| $T$ (K) | $\alpha$ | $l_\varphi$ (nm) |
|---|---|---|
| 2 | -17.3127 | 70.153 |
| 10 | -16.3126 | 79.480 |
| 25 | -12.1401 | 96.281 |
| 50 | -14.0746 | 74.051 |
| 75 | -5.5225 | 92.644 |
| 100 | -6.8147 | 36.384 |
| 110 | 0.4467 | 71.382 |
| 120 | 2.4613 | 83.538 |
| 125 | 5.1379 | 70.626 |

Table S3. The M-F fitting exponents.

| $T$ (K) | $N$ | $l_\varphi$ (nm) | $l_{SO}$ (nm) |
|---|---|---|---|
| 2 | 26.5045 | 87.133 | 27.531 |
| 10 | 33.5420 | 87.806 | 32.191 |
| 25 | 35.9897 | 85.244 | 31.487 |
| 50 | 58.0628 | 61.429 | 31.314 |
| 75 | 36.2129 | 58.171 | 30.556 |
| 100 | 58.5422 | 33.834 | 30.117 |
| 110 | 77.8030 | 23.874 | 30.736 |
| 120 | 45.2608 | 21.190 | 649.234 |
| 125 | 33.9499 | 26.910 | 1336.371 |

**REFERENCES**

(1)     Xu, G.; Wang, W.; Zhang, X.; Du, Y.; Liu, E.; Wang, S.; Wu, G.; Liu, Z.; Zhang, X. X. Weak antilocalization effect and noncentrosymmetric superconductivity in a topologically nontrivial semimetal LuPdBi. *Sci. Rep.* **2014,** 4, (1), 5709.

(2)     Nakamura, H.; Huang, D.; Merz, J.; Khalaf, E.; Ostrovsky, P.; Yaresko, A.; Samal, D.; Takagi, H. Robust weak antilocalization due to spin-orbital entanglement in Dirac material $Sr_3SnO$. *Nat. Commun.* **2020,** 11, (1), 1161.

(3)     Stephen, G. M.; Vail, O. A.; Lu, J.; Beck, W. A.; Taylor, P. J.; Friedman, A. L. Weak Antilocalization and Anisotropic Magnetoresistance as a Probe of Surface States in Topological $Bi_2Te_xSe_{3-x}$ Thin Films. *Sci. Rep.* **2020,** 10, (1), 4845.

(4)     Sun, X.; Li, W.; Wang, X.; Sui, Q.; Zhang, T.; Wang, Z.; Liu, L.; Li, D.; Feng, S.; Zhong, S.; Wang, H.; Bouchiat, V.; Nunez Regueiro, M.; Rougemaille, N.; Coraux, J.; Purbawati, A.; Hadj-Azzem, A.; Wang, Z.; Dong, B.; Wu, X.; Yang, T.; Yu, G.; Wang, B.; Han, Z.; Han, X.; Zhang, Z. Room temperature ferromagnetism in ultra-thin van der Waals crystals of 1T-$CrTe_2$. *Nano Res.* **2020,** 13, (12), 3358-3363.